\documentclass{article}

\usepackage{neurips_2021}

\usepackage[utf8]{inputenc} 
\usepackage[T1]{fontenc}    
\usepackage[french,english]{babel}
\usepackage{hyperref}       
\usepackage{url}            
\usepackage{booktabs}       
\usepackage{amsfonts}       
\usepackage{nicefrac}       
\usepackage{microtype}      

\usepackage{natbib}

\usepackage{amsmath}
\usepackage{amsthm}
\usepackage{amssymb}

\usepackage{algorithm}
\usepackage{algorithmic}
\usepackage[titlenumbered,ruled,noend,algo2e]{algorithm2e}

\SetCommentSty{mycommfont}
\SetEndCharOfAlgoLine{}

\usepackage{stmaryrd} 
\usepackage{booktabs}  
\usepackage{enumitem}

\usepackage{graphicx} 
\graphicspath{{images/}, {prebuiltimages/}, {srcimages/}}
\usepackage{subcaption}

\usepackage{color}
\definecolor{linkcolor}{RGB}{83,83,182}
\definecolor{citecolor}{RGB}{128,0,128}
\hypersetup{
    colorlinks=true,
    citecolor=citecolor,
    linkcolor=linkcolor,
    urlcolor=linkcolor
}

\usepackage[capitalise,nameinlink]{cleveref} 
\usepackage{shortcuts_js}

\Crefname{equation}{Eq.}{Eqs.}
\Crefname{figure}{Fig.}{Figs.}
\Crefname{table}{Tab.}{Tabs.}
\Crefname{algorithm}{Alg.}{Algs.}
\Crefname{appendix}{App.}{Apps.}

\title{
    Electromagnetic neural source imaging under sparsity constraints with SURE-based hyperparameter tuning
}

\author{%
  Pierre-Antoine Bannier \\
  Universit\'e Paris Saclay, Inria, CEA\\
  Palaiseau, 91120, France\\
  \texttt{pierreantoine.bannier@gmail.com} \\
  \And
  Quentin Bertrand \\
  Universit\'e Paris Saclay, Inria, CEA\\
  Palaiseau, 91120, France\\
  \texttt{quentin.bertrand@inria.fr} \\
  \AND
  Joseph Salmon \\
  IMAG, Univ. Montpellier, CNRS \\
  Institut Universitaire de France (IUF) \\
  Montpellier, France \\
  \texttt{joseph.salmon@umontpellier.fr} \\
  \And
  Alexandre Gramfort \\
  Universit\'e Paris Saclay, Inria, CEA\\
  Palaiseau, 91120, France\\
  \texttt{alexandre.gramfort@inria.fr} \\
}

\begin{document}

\maketitle

\vspace{-0.3cm}
\begin{abstract}

Estimators based on non-convex sparsity-promoting penalties were shown to yield state-of-the-art solutions to the magneto-/electroencephalography (M/EEG) brain source localization problem.
In this paper we tackle the model selection problem of these estimators: we propose to use a proxy of the Stein's Unbiased Risk Estimator (SURE) to automatically select their regularization parameters.
The effectiveness of the method is demonstrated on realistic simulations and $30$ subjects from the Cam-CAN dataset.
To our knowledge, this is the first time that sparsity promoting estimators are automatically calibrated at such a scale.
Results show that the proposed SURE approach outperforms cross-validation strategies and state-of-the-art Bayesian statistics methods both computationally and statistically.

\end{abstract}


\section{Introduction}
\label{sec:introduction}
Magneto- and electroencephalography (M/EEG, \cite{Berger1929,Cohen68}) are non-invasive technologies tailored for monitoring electrical brain activity with a high temporal resolution (milliseconds).
Yet, reconstructing the spatial cortical current density at the origin
of M/EEG data is a challenging high-dimensional ill-posed linear inverse problem \citep{Dale_Sereno_1993,Baillet_Mosher_Leahy_2001}.
The M/EEG inverse problem is typically addressed using Lasso-type estimators \citep{Tibshirani96}, and more precisely group-Lasso penalties \citep{Obozinski_Wainwright_Jordan11,Gramfort_Kowalski_Hamalainen12}.
The latter approaches can be refined with non-convex penalties \citep{Candes_Wakin_Boyd08,Strohmeier_Bekhti_Haueisen_Gramfort_2016} that exhibit several advantages: they yield sparser physiologically-plausible solutions, mitigate the intrinsic Lasso amplitude bias, and rely on iterative convex optimization problems which can be solved efficiently with coordinate descent \citep{Tseng_Yun09,Shi_Tu_Xu_Yin16,Massias_Vaiter_Gramfort_Salmon20}.

The major practical bottleneck of these techniques remains the calibration of the regularization parameter, \ie the parameter trading the data-fitting term against the sparsity-promoting prior.
State-of-the-art hyperparameter selection techniques for the M/EEG source localization problem include hierarchical Bayesian modelling
\citep{Molina_Katsaggelos_Mateos1999,Sato_Yoshioka_Kajihara_Toyama_Goda_Doya_Kawato2004} and hyperparameter optimization (HO) \citep{Colson_Marcotte_Savard2007, Kunapuli2008,Franceschi_Frasconi_Salzo_Pontil18}.
Hierarchical Bayesian approaches require to specify a prior distribution on the regularization hyperparameter.
Regression coefficients and the regularization parameter can then
be inferred with multiple techniques \cite{Tipping01,Pereyra_BioucasDias_Figueiredo_2015,Bekhti_Badeau_Gramfort_2017,Vidal_Bortoli_Pereyra_Durmus_2020}.
The idea of HO is to select the regularization parameter such that the regression coefficients minimize a given criterion.
This then boils down to a bilevel optimization problem can then be solved using zero-order methods \citep{Bergstra_Bengio12,Li_Jamieson_DeSalvo_Rostamizadeh_Talwalkar17,Akiba_Sano_Yanase_Ohta_Koyama2019} or first-order methods \citep{Franceschi_Donini_Frasconi_Pontil17,Bertrand_Klopfenstein_Blondel_Vaiter_Gramfort_Salmon2020,Bertrand_Klopfenstein2021}.
Popular statistical criteria include K-fold cross-validation (folds are created across sensors \cite{Craven_Wahba_79}), which is however not well-suited for the M/EEG inverse problem: samples are not \iid due to spatially-correlated sensors \citep{Dale_Sereno_1993}.
Therefore, the crucial question revolves around finding a criterion to properly identify the neural generators at the origin of the observed signal. Stein \cite{Stein81} proposed an unbiased estimator of the quadratic risk of estimators: Stein's Unbiased Risk Estimator (SURE), which has proved to be well-suited for inverse problem hyperparameter selection \citep{Blu_Luisier07,Ramani_Blu_Unser08,Pesquet_Benazza_Chaux_2009}.

In this paper, we combine reweighting techniques with a SURE-based model selection to identify active brain sources.
The main contributions of this paper are as follows:
\begin{enumerate}[nolistsep]
    \item We combined the SURE criterion with a non-convex estimator.
    This yields a fast and parameter-free approach to select the regularization parameter for the \texttt{irMxNE} algorithm.
    \item With experiments on more than $30$ subjects from the Cam-CAN biobank \citep{Camcan}, we extensively show that the proposed SURE-based hyperparameter selection technique for \texttt{irMxNE} achieves state-of-the-art results on real M/EEG data.
    \item Code is available at \url{https://github.com/PABannier/automatic_hp_selection_for_meg} and already disseminated via the popular brain imaging package \texttt{MNE} \citep{mne}.
\end{enumerate}
\textbf{Notation.}
The Frobenius norm of $\mathbf{A}$ is denoted by $\normin{\mathbf{A}}_{\text{F}}$.
For any integer $d \in \mathbb{N}$, we denote by $[d]$ the set $\{1, \dots, d\}$.
With M/EEG data, $N$ is the
number of sensors, $T$ the number of time instants of the measurements, and $S$ the number of source locations positioned on the cortical mantle.

%
\section{Method}
\label{sec:sure}
\textbf{Model.}
The M/EEG source localization problem can be cast as a high-dimensional inverse problem:
\begin{equation} \label{eq:1}
    \mathbf{M = GX^* + E} \enspace,
\end{equation}
where $\mathbf{M} \in \mathbb{R}^{N \times T}$ is a measurement matrix,
$\mathbf{G} \in \mathbb{R}^{N \times P}$ is the (known) design matrix,
$\bfX^* \in \bbR^{P \times T}$ is the unknown true regression parameters,
$\mathbf{E} \in \mathbb{R}^{N \times T}$ is the noise matrix, which is assumed Gaussian \iid: $\mathbf{E}_{i, t} \sim \mathcal{N}(0, 1), \forall i \in [N], \forall t \in [T]$.
The coefficient matrix $\mathbf{X^*} \in \mathbb{R}^{P \times T}$ is to be recovered from the observation of $\mathbf{M}$ and $\mathbf{G}$.
While simpler models assume the source orientations are normal to the cortical mesh, we rely on \emph{free orientation} models, where amplitudes and orientations are jointly inferred.
It boils down to reconstructing the amplitudes of three orthogonal sources at each spatial location: $P=3S$ and $\bfX \in \bbR^{3S \times T}$ is partitioned into $S$ blocks $\bfX_s \in \mathbb{R}^{3 \times T}$, where each $\bfX_s$ is the block corresponding to the $\mathrm{s^{th}}$ source location \citep{Gramfort_Strohmeier_Haueisen_Hamalainen_Kowalski13}.

\textbf{Estimator.}
Given a regularization parameter $\lambda > 0$, the \texttt{irMxNE} optimization problem reads:
\begin{equation} \label{eq:reweighted_lasso}
    \widehat{\bfX}^{(\lambda)}
    \in
    \argmin_{\bfX \in \bbR^{P \times T}}
    \frac{1}{2}\normin{\mathbf{M - GX}}^2_{\text{F}}
    + \lambda
    \sum_{s=1}^S \sqrt{\normin{\mathbf{X}_s}_{\mathrm{F}}}
    \enspace .
\end{equation}
Non-convex problem (\ref{eq:reweighted_lasso}) is solved by iterativaly solving convex problems \cite{Candes_Wakin_Boyd08, Strohmeier_Bekhti_Haueisen_Gramfort_2016}, see \Cref{alg:irmxne}.

\textbf{Model selection.}
The parameter $\lambda$ is often chosen such that the corresponding regression coefficients
$\widehat{\bfX}^{(\lambda)}$ (in \Cref{eq:reweighted_lasso}) minimize a given criterion.
\cite{Stein81} proposed a criterion to avoid overcomplex models:
\begin{align} \label{eq:7}
    \text{SURE}(\widehat{\bfX}^{(\lambda)})
    &= -NT\sigma^2 + \normin{
        \mathbf{M} - \mathbf{G} \widehat{\bfX}^{(\lambda)}
    }_{\text{F}}^2
    + 2 \sigma^2 \text{dof}(\widehat{\bfX}^{(\lambda)})
    \enspace ,
\end{align}
where $\text{dof}(\widehat{\bfX})$ is the degrees of freedom of $\widehat{\bfX}$ \cite{Efron86} and $\sigma$ is the true noise level of the measurements.
The non-convex penalty used in \Cref{eq:reweighted_lasso} prevents the derivation of a closed-form formula for the \texttt{irMxNE} degree of freedom.
To circumvent this issue, multiple SURE proxies have been proposed \citep{Ramani_Blu_Unser08}.
In this work we use the Finite-Difference Monte-Carlo SURE (FDMC SURE, \cite{Deledalle_Vaiter_Fadili_Peyre14}, \Cref{alg:compute_sure_mcfd}).
Due to the non-convexity of the problem, the $\text{dof}$ term can only be evaluated numerically \citep{Mazumder_Friedman_Hastie11}.


\section{Experiments}
\label{sec:experiments}
%
Several strategies have been proposed to calibrate the regularization parameter in \Cref{eq:reweighted_lasso}:
SURE (proposed, \Cref{alg:compute_sure_mcfd} in \Cref{app:sub_sure}),
spatial cross-validation \citep{Stone_Ramer65},
and hierarchical Bayesian model with a Gamma hyperprior on $\lambda$ ($\lambda$-MAP, \citep{Bekhti_Badeau_Gramfort_2017}).
It is worth mentioning that $\lambda$-MAP is not a fully automatic method: the default value of the Gamma hyperprior parameter $\beta$ given in \citep{Bekhti_Lucka_Salmon_Gramfort17} yields poor results in our experiments.
We had to handtune $\beta=10$ for simulated data, and $\beta=5$ for real data.
Data is assumed to be whitened, therefore the true noise level $\sigma$ is assumed to be known and equal to $1$~\citep{Engemann_gramfort_2015}.

\begin{figure}[tb]
    \begin{subfigure}{1\textwidth}
      \begin{center}
        \includegraphics[width=.5\textwidth]{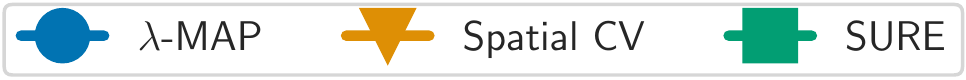}
      \end{center}
    \end{subfigure}

  \begin{subfigure}{1\textwidth}
      \includegraphics[width=\textwidth]{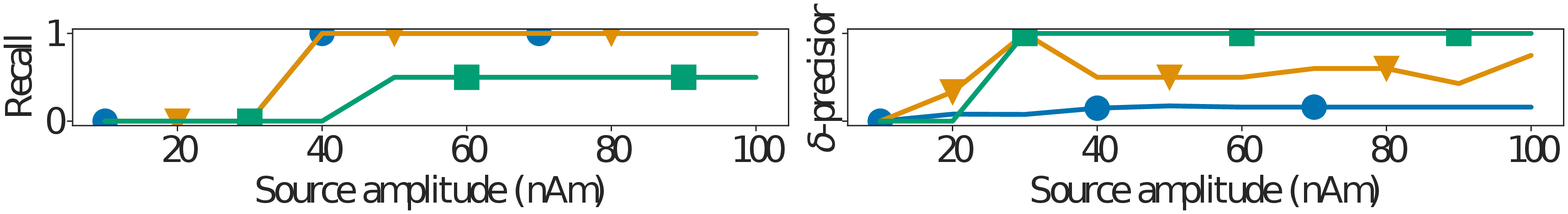}
  \end{subfigure}
    \caption[placeholder]{\textbf{Simulated data, statistics on the active set recovery.}
    Statistics for each model selection procedure on the reweighted Lasso, fitted on a simulated setup from the left auditory task.
    }
    \label{fig:simulated_data}
\end{figure}

\textbf{Experiments on simulated data (\Cref{fig:simulated_data}).}
We compared the robustness of each hyperparameter selection technique in various signal-to-noise ratio regimes.
We simulated two sources, one in each auditory cortex and varied their amplitude.
We computed $\delta$-statistics \citep{Chevalier_Salmon_Thirion_18} on the recovery of the active set, and chose an extent with 7\,mm of radius.
SURE correctly reconstructs the active sources (1 of $\delta$-precision) but fails at identifying all of them.
Since both spatial CV and $\lambda$-MAP tend to overfit the data, they predict larger supports than expected and yield high recall values.

\textbf{Experiments on real data (\Cref{tab:camcan_res,fig:real_data_camcam}).}
We compared hyperparameter selection for the regularization parameter $\lambda$ of the reweighted multitask Lasso  (\Cref{eq:reweighted_lasso}) on 30 subjects of the Cam-CAN biobank dataset \citep{Camcan}.
Data consists in the recording of $N=306$ magnetometers and gradiometers after a left auditory stimulation ($T=71$ time samples).
We used \Cref{eq:reweighted_lasso} to estimate the active sources among the $P=24,582$ source candidates.
For each hyperparameter selection technique, we obtain active sources for each subject that we represent on an average brain in \Cref{fig:real_data_camcam}.
Summary statistics of the experiment are provided in \Cref{tab:camcan_res}.
Regarding computational efficiency, $\lambda$-MAP, spatial CV and SURE took respectively $8$, $1184$ and $492$ seconds per subproblem for the 30 subjects.

\Cref{tab:camcan_res} shows that spatial CV consistently overfits the dataset by choosing a too small regularization hyperparameter. As predicted by theory \citep{Shao_1993, Arlot_Celisse10}, spatial CV yields overly large supports, most often with more than two sources.
$\lambda$-MAP provides an average number of recovered sources similar to SURE.
Nonetheless, SURE yields a significantly better explained variance.

\Cref{fig:real_data_camcam} shows the reconstructed sources after a left auditory stimulation on $30$ subjects, registered on an average template brain.
Each dot represents an active source in the brain, each color corresponds to a subject.
The areas colored in red are the auditory cortices obtained using a functional atlas \cite{freesurfer}.
Ideally, we expect one reconstructed source in each auditory cortex, and a low dispersion of the sources across subjects.
Spatial CV yields sources all over the brain surface.
$\lambda$-MAP correctly localizes the sources but fails to recover any sources on almost one third of the subjects (see \Cref{tab:camcan_res}).
SURE always recovers correctly at least one source,
and often the correct two sources in both cortices.
\def\scale{0.13}
\def\surescale{0.26}
\def\sizesubfig{0.32}
\begin{figure}[tb]
    \begin{subfigure}{\sizesubfig\textwidth}
        \includegraphics[width=\textwidth]{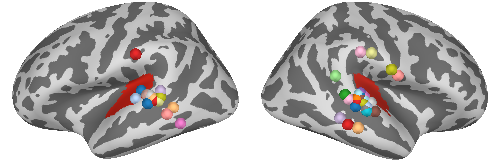}
      \caption{SURE.}
      \label{fig:sure}
    \end{subfigure}
    \hfill
    \begin{subfigure}{\sizesubfig\textwidth}
        \includegraphics[width=\textwidth]{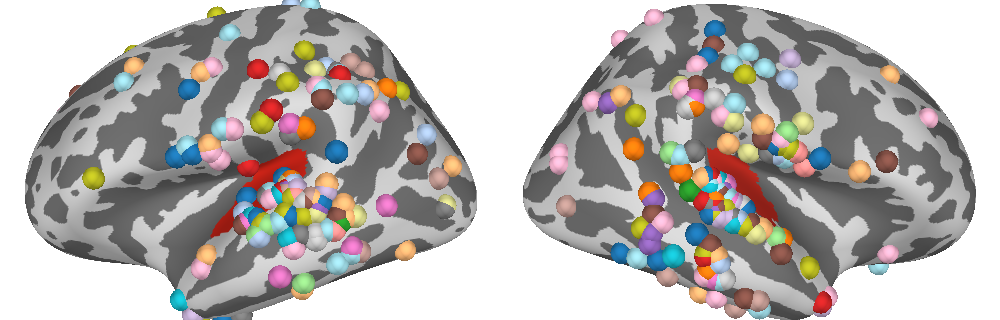}
      \caption{Spatial CV.}
      \label{fig:spatial_cv}
    \end{subfigure}
    \hfill
    \begin{subfigure}{\sizesubfig\textwidth}
        \includegraphics[width=\textwidth]{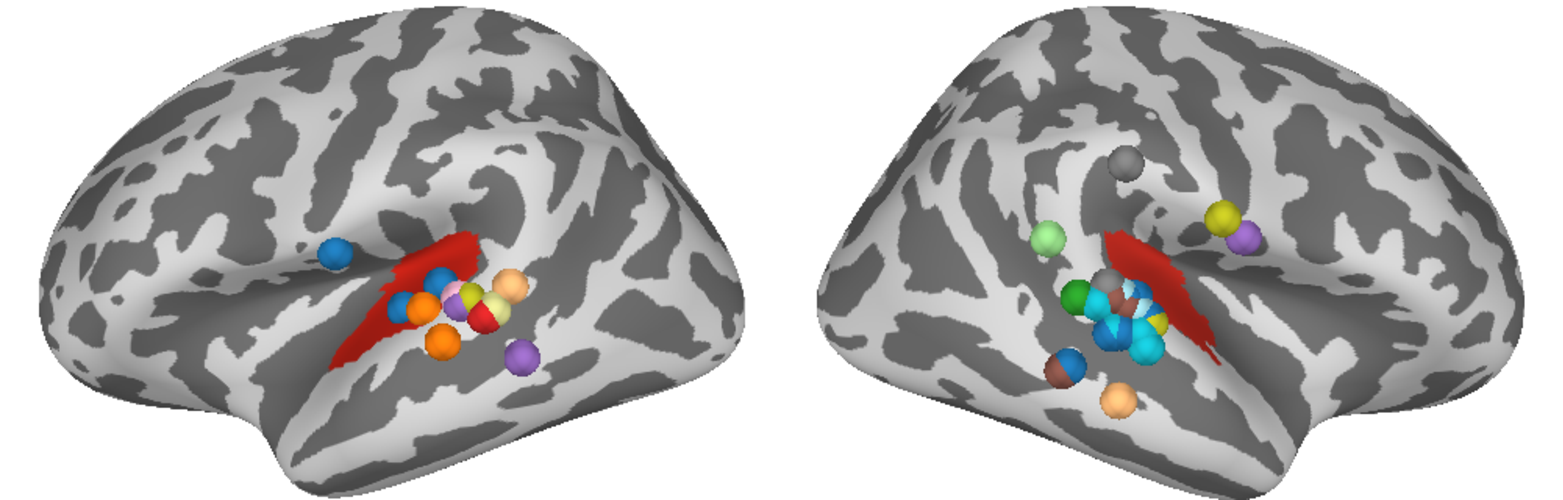}
        \caption{$\lambda$-MAP.}
      \label{fig:hbm}
    \end{subfigure}
    \caption[placeholder]{\textbf{Real data, brain source locations aggregated across subjects.}
    Brain source locations reconstructed in each hemisphere after an auditory stimulation for each model selection procedure.
    }
    \label{fig:real_data_camcam}
\end{figure}
\begin{table}
\caption{Real data, aggregated results.}
\label{tab:camcan_res}
\begin{tabular}{ p{3cm}|p{3cm} p{3cm} p{3cm}  }
 Average metrics & Spatial CV & SURE & $\lambda$-MAP\\
 \hline
 $\lambda / \lambda_{\text{max}}$   & 0.30 & 0.61 & 0.9 \\
 Explained variance &   \textbf{0.67}  & 0.32 & 0.06 \\
 \# of sources & 9.30 & \textbf{1.44} & 1.3 \\
 \hline
 \% of zero sources    & 0 & \textbf{0} & 29.63\\
 \% of one source &   0  & \textbf{55.56} & 40.74\\
 \% of two sources &   3.7  & \textbf{44.44} & 18.52\\
 \% of $> 2$ sources & 96.3  & \textbf{0} & 11.11\\
\end{tabular}
\end{table}

\clearpage

\section{Broader impact}
This paper paves the way to a wider adoption of recent results in machine learning in the context of non-invasive brain source imaging commonly employed in cognitive and clinical neuroscience.
Since the code is already available in the \texttt{MNE} package,
the proposed algorithm could soon be impactful in the neuroimaging community.

\section*{Acknowledgments}
This work was supported by the ERC-StG-676943 SLAB.

\bibliographystyle{plainnat} 
\bibliography{references_all} 

\clearpage
\appendix


\section{\texttt{MxNE}}
\label{app:mxne_details}

We recall that $\bfG \in \bbR^{N \times P}$ denotes the design matrix, $\bfM \in \bbR^{N \times T}$ is the measurement matrix.
Given a regularization parameter $\lambda > 0$, the \texttt{MxNE} optimization problem \citep{Gramfort_Kowalski_Hamalainen12} reads:
\begin{equation} \label{eq:mxne_optim_pb}
    \widehat{\bfX}^{(\lambda)}
    \in
    \argmin_{\bfX \in \bbR^{P \times T}}
    \frac{1}{2}\normin{\mathbf{M - GX}}^2_{\text{F}}
    + \lambda
    \sum_{s=1}^S \normin{\mathbf{X}_s}_{\mathrm{F}}
    \enspace .
\end{equation}
It relies on a convex optimization problem that can be solved using block coordinate descent solvers \citep{Tseng_Yun09,Shi_Tu_Xu_Yin16, Massias_Vaiter_Gramfort_Salmon20}.

\section{Algorithms details}
\label{app:algo_details}

\subsection{FDMC SURE}
\label{app:sub_sure}
%

Evaluating the FDMC SURE consists in solving the following bilevel optimization problem \citep{Deledalle_Vaiter_Fadili_Peyre14}:
\begin{align}
\begin{split}
\label{eq:bilevel_sure_mcfd}
    &\hat{\lambda} \in \argmin_{\lambda \in \bbR}
    \frac{1}{2} \normin{
        \bfM - \bfG\widehat{\bfX}^{(\lambda, 1)}
    }^2_{\text{F}}
    + \frac{2\sigma^2}{\epsilon} \langle \bfG
    (\widehat{\bfX}^{(\lambda, 2)} - \widehat{\bfX}^{(\lambda, 1)}),
    \mathbf{\Delta} \rangle \\
    \text{s.t.} \enspace &\widehat{\bfX}^{(\lambda, 1)}
    \in \argmin_{\bfX \in \bbR^{P \times T}}
    \frac{1}{2} \normin{
        \bfM - \mathbf{GX}
    }^2_{\text{F}}
    + \lambda
    \sum_{s=1}^S \sqrt{\normin{\bfX_s}_{\mathrm{F}}}
    \\
    &\widehat{\bfX}^{(\lambda, 2)}
    \in \argmin_{\bfX \in \bbR^{P \times T}}
    \frac{1}{2} \normin{
        \bfM + \epsilon \mathbf{\Delta} - \mathbf{GX}
    }^2_{\text{F}}
    + \lambda
    \sum_{s=1}^S \sqrt{\normin{\bfX_s}_{\mathrm{F}}}
    \enspace ,
\end{split}
\end{align}
with $\epsilon > 0$ and
$\mathbf{\Delta} \in \bbR^{N \times T}$ a matrix which coefficients
are independent and identically distributed from a normal distribution of mean $0$ and of variance $1$.
Since data is assumed to be whitened, $\sigma$ is set to $1$.
The finite difference step $\epsilon$ is chosen using the heuristic from \cite{Deledalle_Vaiter_Fadili_Peyre14}:
$\epsilon=\frac{2\sigma}{N^{0.3}}$.
\Cref{alg:irmxne} shows how to solve the inner problem of \Cref{eq:bilevel_sure_mcfd}.
The computation of the FDMC SURE can be found in
\Cref{alg:compute_sure_mcfd}. Below $\odot$ stands for the element-wise multiplication.
\begin{minipage}[t]{\linewidth}
    \begin{minipage}[t]{0.49\linewidth}
        {\fontsize{4}{4}\selectfont
        \begin{algorithm}[H]  
        \SetKwInOut{Input}{input}
        \SetKwInOut{Init}{init}
        \SetKwInOut{Parameter}{param}
        \caption{\textsc{\texttt{irMxNE}} \citep{Strohmeier_Bekhti_Haueisen_Gramfort_2016}
        }
        \label{alg:irmxne}
        \Input{$
            \bfG \in \bbR^{N \times P},
            \bfM \in \bbR^{N\times T}, \lambda > 0, K \in \bbN$
            }
        \Init{$
            \widetilde{\bfX} = \mathbf{0}_{\bbR^{P \times T}},
            \bfW = \bfI_P,
            \bfw = \mathbf{0}_S$ \\
            $\epsilon = 10^{-8}$}
    
        \For{$k = 1, \dots, K$}{
            $\widetilde \bfG = \bfG\bfW$
    
            \tcp{Iteratively solve convex problems}
            $\widetilde{\bfX} \leftarrow$ $\texttt{MxNE}(\widetilde \bfG, \bfM)$
            \tcp*[r]{Using $\widetilde \bfX$ to warm start}
    
            $\bfX = \bfW\widetilde{\bfX}$
    
            $\bfw = ((2\normin{\bfX_{s}}_{\text{F}} + \epsilon)^{-1})_{s \in [S]}$
    
            $\bfW = \text{diag}(\bfw \otimes \mathbf{1}_{(3)})$
        }
    
        \Return{$\bfX$}
        \label{alg:ixMxNE}
        \end{algorithm}
        }
    \end{minipage}
    \hfill
    \begin{minipage}[t]{0.49 \linewidth}
        {\fontsize{4}{4}\selectfont
        \begin{algorithm}[H]  
        \SetKwInOut{Input}{input}
        \SetKwInOut{Init}{init}
        \SetKwInOut{Parameter}{param}
        \caption{
            \textsc{Compute FDMC SURE} \\
            (adapted from \cite{Deledalle_Vaiter_Fadili_Peyre14})
        }
        \label{alg:compute_sure_mcfd}
        \Input{$
            \bfG \in \bbR^{N \times P},
            \bfM \in \bbR^{N\times T}, \lambda > 0, K \in \bbN$}
        \Init{$
            \mathbf{\Delta} \in \bbR^{N\times T},
            \epsilon > 0,
            \sigma > 0$\\
            $\forall (n, t) \in [n] \times [T], \mathbf{\Delta}_{n, t} \sim \cN(0, 1)$
            }
    
            $\bfX^{(\lambda, 1)}
            \leftarrow$ \Cref{alg:ixMxNE}$(\bfG, \bfM, \lambda, K)$
    
            $\bfX^{(\lambda, 2)}
            \leftarrow$ \Cref{alg:ixMxNE}$(\bfG, \bfM + \epsilon \Delta, \lambda, K)$
    
            \tcp{Degree of freedom computation}
    
            $\mathbf{J} = \bfG (\bfX^{(\lambda, 2)} - \bfX^{(\lambda, 1)}) \odot \mathbf{\Delta}$
    
            \tcp{Finite-difference}
            $\text{dof} = \frac{1}{\epsilon} \sum_{i=1}^N \sum_{j=1}^T \mathbf{J}_{i,j}$
    
            \tcp{FDMC SURE computation}
            $\mathrm{SURE} =
            \normin{
                \bfM - \bfG\bfX^{(\lambda, 1)}
                }^2_\text{F}
            - NT\sigma^2 + 2\sigma^2\text{dof}$
    
        \Return{$\mathrm{SURE}$}
        \end{algorithm}
        }
    \end{minipage}
\end{minipage}

\vspace{0.2in}

Combining both algorithms, we propose in \Cref{alg:grid_search_sure_mcfd} the procedure to automatically select $\lambda > 0$ for \texttt{irMxNE}.

    {\fontsize{4}{4}\selectfont
    \begin{algorithm}[H]  
    \SetKwInOut{Input}{input}
    \SetKwInOut{Init}{init}
    \SetKwInOut{Parameter}{param}
    \caption{
        \textsc{Grid-search to solve Problem \ref{eq:bilevel_sure_mcfd}}
    }
    \label{alg:grid_search_sure_mcfd}
    \Input{$
        \bfG \in \bbR^{N \times P},
        \bfM \in \bbR^{N \times T},
        \lambda_1, \dots, \lambda_n > 0,
        K \in \bbN$
        }
    \Init{$
        \mathrm{SURE}_{\mathrm{opt}} = +\infty$}

    \For{$\lambda=\lambda_1, \dots, \lambda_n$} {
            $\mathrm{SURE} \leftarrow $\Cref{alg:compute_sure_mcfd}$(\bfG, \bfM, \lambda, K)$

            \If{$\mathrm{SURE} < \mathrm{SURE}_{\mathrm{opt}}$}{
                $\lambda_{\mathrm{opt}} = \lambda$; $\mathrm{SURE}_{\mathrm{opt}}= \mathrm{SURE}$
            }
    }

    \Return{$\lambda_{\mathrm{opt}}$}
    \end{algorithm}
    }

\subsection{Cross-validation}

Cross-validation consists in partitioning data $(\bfG, \bfM)$ into $V \in \mathbb{N}^*$ hold-out datasets $(\bfG^{\text{train}_v}, \bfM^{\text{train}_v})$, $v\in [V]$. The matrices are partitioned along the row axis, a setup referred to as spatial cross-validation when dealing with M/EEG data. The regularization parameter $\lambda$ is then chosen to minimize the averaged squared norm of the errors:
\begin{align}
\begin{split}
    &\hat{\lambda} \in \argmin_{\lambda \in \bbR}
    \frac{1}{V} \sum_{v=1}^{V}
    \normin{
        \bfM^{\text{val}_v} - \bfG^{\text{val}_v}\widehat{\bfX}^{(\lambda, v)}
    }^2_{\text{F}} \\
    \text{s.t.} \enspace
    &\widehat{\bfX}^{(\lambda, v)}
    \in \argmin_{\bfX \in \bbR^{P \times T}}
    \frac{1}{2} \normin{
        \bfM^{\text{train}_v} - \mathbf{G^{\text{train}_v}X}
    }^2_{\text{F}}
    + \lambda
    \sum_{s=1}^S \sqrt{\normin{\bfX_s}_{\mathrm{F}}}
    \quad ,
    \quad
    \forall v \in [V]
    \enspace .
\end{split}
\end{align}






\subsection{Hierarchical Bayesian modelling}

\vskip 0.1in

{\fontsize{4}{4}\selectfont
\begin{algorithm}[h]  
\SetKwInOut{Input}{input}
\SetKwInOut{Init}{init}
\SetKwInOut{Parameter}{param}
\caption{\textsc{$\lambda$-MAP \citep{Bekhti_Badeau_Gramfort_2017}}
\label{alg:lambda_map}
}
\Input{$
    \bfG \in \bbR^{N \times P},
    \bfM \in \bbR^{N \times T},
    \lambda^{(0)} \in \bbR^+,
    n_{\mathrm{iter}} \in \bbN^*,
    \beta > 0,
    \epsilon > 0
    $}
\Init{$
    \lambda_{\max} = \normin{\bfG^{\top}\bfM}_{2, \infty}
    $,
    $m = \lambda_{\max} / 2$,
    $\alpha = m \beta + 1$
    }

    \For{$i = 1,\dots, n_{\mathrm{iter}}$}
    {
        $\bfX \leftarrow$ \Cref{alg:ixMxNE} (
            $\bfG$, $\bfM$, $\lambda^{(i-1)}$)
            \tcp{Solve \texttt{ixMxNE} problem}

        $\lambda^{(i)} = (2ST + \alpha - 1) / (\sum_{s=1}^S \sqrt{\normin{\bfX_s}_{\mathrm{F}}} + \beta)$

        \If{$\lvert \lambda^{(i)} - \lambda^{(i-1)} \rvert < \epsilon$}
        {
          \Return{$\lambda^{(i)}$}
        }
    }

\Return{$\lambda^{n_\mathrm{iter}}$}
\end{algorithm}
}

For $\lambda$-MAP, we had to fine-tune by hand the hyperprior parameter. We set $\beta=10$ for simulated data, and $\beta=5$ for real data. While $\lambda$-MAP is robust to the initialization of $\lambda$, it remains highly dependent of $\beta$, making it not fully automatic. Some experiments required an order of magnitude larger $\beta$ to prevent the iterate scheme to converge over $\lambda_{\text{max}}$. Indeed, we have noticed that $\lambda$-MAP often selects a hyperparameter $\lambda$ larger than $\lambda_{\text{max}}$, due to a poorly-chosen $\beta$.

\subsection{Warm start}
%
To accelerate the grid-search procedure, we sequentially solve the first convex subproblems (before the first reweighting) and initialize the weights of the $i$-th Lasso estimator $\widehat{\bfX}^{(\lambda_{i})}$ with $\widehat{\bfX}^{(\lambda_{i-1})}$.
This computational trick is known as warm start (\Cref{alg:grid_search_warm_start}).
{\fontsize{4}{4}\selectfont
\begin{algorithm}[h]  
\SetKwInOut{Input}{input}
\SetKwInOut{Init}{init}
\SetKwInOut{Parameter}{param}
\caption{\textsc{Efficient warm start for reweighted Lasso with grid-search}
\label{alg:grid_search_warm_start}
}
\Input{$
    \bfG \in \bbR^{N \times P},
    \bfM \in \bbR^{N \times T},
    \lambda_1, \dots, \lambda_n > 0,
    K \in \bbN^*$}
\Init{$\bfX^{(\lambda_{0})} = \textbf{0}_{N \times T}$}
    \tcp{First solve problems without reweighting}

    \For{$i \in [n]$}
    {
        \tcp{This leads a more efficient warm start}
        $\bfX^{(\lambda_i)}
        \leftarrow
        \texttt{MxNE}(\bfG, \bfM, \lambda_i)$
        \tcp*[l]{Solve \texttt{MxNE} using $\bfX^{(\lambda_{i-1})} $ to warm start}
    }

    \tcp{Then solve the remaining subproblems}
    \For{$i \in [n]$}
    {
        $\bfX^{(\lambda_i)}
        \leftarrow
        \texttt{irMxNE}(\bfG, \bfM, \lambda_i, K-1)$
        \tcp*[l]{Solve \texttt{irMxNE} using $\bfX^{(\lambda_{i})} $ to warm start}
    }

\Return{$\bfX^{(\lambda_1)}, \dots, \bfX^{(\lambda_n)}$}
\end{algorithm}
}

\end{document}